\documentclass[twocolumn,showpacs,preprintnumbers,amsmath,amssymb]{revtex4}

\usepackage{graphicx}
\usepackage{dcolumn}
\usepackage{amssymb}

\begin{document}


\title{Formation and relaxation of RbHe exciplexes on He nanodroplets studied by femtosecond pump and picosecond probe spectroscopy}

\author{C. Giese$^1$}
\author{T. Mullins$^{2*}$}
\author{B. Gr{\"u}ner$^1$}
\author{M. Weidem{\"u}ller$^2$}
\author{F. Stienkemeier$^1$}
\author{M. Mudrich$^1$}
\affiliation{$^1$Physikalisches Institut, Universit\"at Freiburg, 79104 Freiburg, Germany}
\affiliation{$^2$Physikalisches Institut, Universit\"at Heidelberg, 69120 Heidelberg, Germany}

\date{\today}

\begin{abstract}
Vibrationally resolved photoionization spectra of RbHe exciplexes forming on He nanodroplets are recorded using femtosecond pump-probe spectroscopy with amplitude-shaped probe pulses. The time-evolution of the spectra reveals an exciplex formation time $\sim 10\,$ps followed by vibrational relaxation extending up to $\gtrsim 1\,$ns. This points to an indirect, time-delayed desorption process of RbHe off the He surface.
\end{abstract}

\maketitle

\section{\label{sec:Intro}Introduction}
The success of the rapidly developing field of spectroscopy of doped helium (He) nanodroplets is based on applying ever new laser techniques as they become available. Such studies are motivated on the one hand by the unique properties of He nanodroplets which make them a nearly ideal spectroscopic matrix~\cite{Toennies:2004,Stienkemeier:2006}. The small size, low temperature, the superfluid state as well as the inertness with respect to light absorption and to chemical reactivity makes He nanodroplets a suitable environment for high resolution spectroscopic studies of isolated atoms, molecules, aggregates and clusters. On the other hand, the guest-host interactions between the dopant atoms or molecules and the He droplets are being studied for their own interest, in particular for probing superfluidity on the nanoscale.

He droplets constitute indeed an extremely weakly perturbing environment as far as dopant molecules in their electronic ground state are concerned. However, upon electronic excitation or ionization of the dopants, strong interactions with the He may induce considerable spectroscopic shifts and broadenings~\cite{Stienkemeier2:1995,Buenermann:2007,Pentlehner:2010,Loginov:2005}. The He then turns from an inert substrate into an active reaction partner which facilitates the formation of He containing molecules ('exciplexes')~\cite{Reho:1997,Reho:2001,Reho2:2001,Bruehl:2001,Schulz:2001,Droppelmann:2004,Mudrich:2008}, ionic complexes ('snowballs')~\cite{DoeppnerJCP:2007,Mueller:2009,Theisen:2010} and even nanoplasmas~\cite{MikaberidzePRL:2009,DoeppnerPRL:2010,KrishnanPRL:2011,Krishnan:2012}.

Exciplexes consisting of one excited alkali (Ak) metal atom Ak$^*$  and one or a few He atoms formed on He nanodroplets have been extensively studied spectroscopically and more recently by means of photoionization mass spectrometry and ion imaging techniques~\cite{Stienkemeier:1996,Reho:1997,Reho:2000,Bruehl:2001,Schulz:2001,Droppelmann:2004,Mudrich:2008,LoginovPhD:2008,LoginovJPCA:2011,Fechner:2012}. The structure and spectra of Ak$^*$He exciplexes have been investigated theoretically using various techniques~\cite{Stienkemeier:1996,Takayanagi:2004,Pacheco:2007,Leino:2011,Chattopadhyay:2012}. While the equilibrium properties of exciplexes are now well characterized even including an environment given by a He cluster or film, the dynamics of the formation process of Ak$^*$He exciplexes still eludes from an accurate description.

One reason for the difficulty of measuring exciplex formation times is related to the location of Ak dopants on the surface of He nanodroplets in shallow dimple states~\cite{Dalfovo:1994,Ancilotto:1995,Stienkemeier2:1995}. Upon laser excitation, the excited Ak$^*$ generally tends to desorb off the droplet surface due to repulsive forces acting between the Ak$^*$ atom and the He droplet as a whole~\cite{Stienkemeier:1996,Buenermann:2007,Callegari:2011}. Thus, in the situation of attractive Ak$^*$-He pair interactions the dynamics initiated by laser absorption is likely to be determined by the competition between the dissociation of the Ak$^*$He$_N$ complex and the formation of the Ak$^*$He$_n$ exciplex molecules, where $n=1,2,\dots$ and He$_N$ denotes the He droplet. For the case of rubidium exciplexes Rb$^*$He formed in the second excited electronic state $1^2\Pi_{3/2}$, though, a weakly bound configuration on the He surface was predicted~\cite{Leino:2011}. The fact that in the experiment Rb$^*$He exciplexes have been observed as desorbed free molecules was rationalized by droplet-induced vibrational relaxation into lower vibrational levels thereby liberating enough energy for evaporating the Rb$^*$He off the droplet surface~\cite{Leino:2011}.

The dynamics of the formation of sodium and potassium exciplexes (Na$^*$He, K$^*$He) has first been studied by time-resolved emission spectroscopy, yielding formation times in the range of tens of ps~\cite{Reho2:2000}. More recently, the K$^*$He and Rb$^*$He exciplex formation dynamics has been probed using femtosecond (fs) pump-probe techniques which revealed a Rb$^*$He signal rise time of 8.5\,ps~\cite{Schulz:2001,Droppelmann:2004,Mudrich:2008}. Theoretical models including one-dimensional semiclassical tunneling~\cite{Reho2:2000}, quantum-classical modeling~\cite{Pacheco:2007}, semiclassical path integral molecular dynamics~\cite{Takayanagi:2004}, and quantum Monte Carlo approaches~\cite{Leino:2011} have predicted values for the exciplex formation times ranging from 1.7\,ps for lithium-helium Li$^*$He, to 31\,ps for Rb$^*$He in the $1^2\Pi_{3/2}$ state.

Recently, new insights have been gained from ion imaging experiments with He droplets doped with Na. Depending on the electronic state to which the NaHe$_N$ complex is excited, qualitatively different velocity distributions of the desorbing Na$^*$He and Na$^*$He$_2$ exciplexes as compared to the neat Na atom were observed~\cite{LoginovPhD:2008}. The isotropic velocity distribution of Na$^*$He measured upon excitation to the $1^2\Pi_{3/2}$ state points to an indirect desorption process which is possibly driven by droplet-induced vibrational relaxation and thus akin to statistical evaporation.

In the present work we study the formation of Rb$^*$He exciplexes initiated by exciting Rb-doped He droplets to the $1^2\Pi_{3/2}$ state of the Rb$^*$He$_N$ complex by means of a new time-resolved spectroscopic approach. Using a fs pulse shaper for amplitude shaping one of the two identical fs pulses we realize a pump-probe scheme where the first pump pulse is ultrashort (200\,fs) and broadband (75\,cm$^{-1}$) and the time delayed probe pulse is stretched to $3.4\,$ps using a pulse shaper which acts as a tunable spectral band-pass filter with a band width of $9.1\,$cm$^{-1}$. In this way, we can resolve the vibrational spectrum of Rb$^*$He and follow its evolution on the picosecond (ps) up to nanosecond (ns) time scales. Surprisingly, we find that a redistribution of the spectral line intensity proceeds as long as 1.7\,ns after the pump pulse. This points to the Rb$^*$He exciplexes being formed within a few ps and then remaining at least partly attached to the droplet surface and slowly relaxing into low vibrational states during hundreds of ps. A similar relaxation dynamics was previously observed with Rb$_2$, Rb$_3$, Na$_3$, and K$_3$ attached to He droplets~\cite{Gruner:2011,Giese:2011,Reho:2001}.

\section{Experimental setup}
\label{sec:Setup}
\begin{figure}
\begin{center}{
\includegraphics[width=0.45\textwidth]{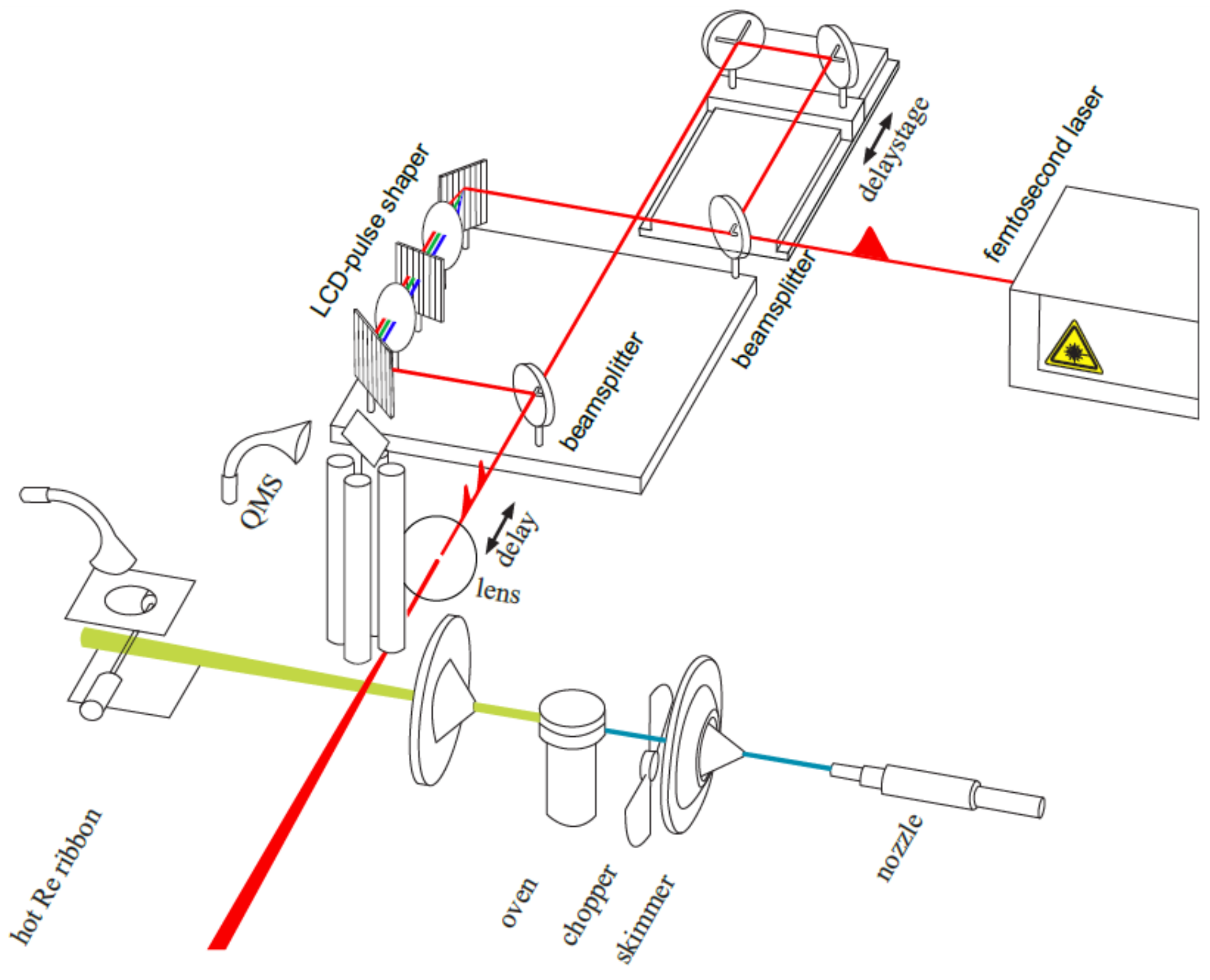}}
\caption{Sketch of the experimental setup. The upper part shows the femtosecond laser and optics including the pulse shaper placed in one arm of the interferometer. On the bottom part the helium droplet beam line is schematically represented.}
\label{fig:setup}
\end{center}
\end{figure}
The experimental arrangement used to produce a beam of Rb-doped helium nanodroplets and to detect photoions created by femtosecond laser ionization is schematically represented in Fig.~\ref{fig:setup} and is similar to previously used setups~\cite{Mudrich:2009,Gruner:2011,Giese:2011}. In short, ultrapure He gas is expanded at high pressure ($p=50\,$bar) through a cold nozzle ($T=20\,$K, diameter $d=5\,\mu$m) into vacuum. At these expansion conditions the average size of the He droplets amounts to $\left\langle N\right\rangle\approx 4000$ He atoms~\cite{Toennies:2004,Stienkemeier:2006}. The droplets enter the adjacent doping chamber through a skimmer ($d=400\mu$m) where they pick up single Rb atoms on their way through a stainless steel pickup cell with a length of 1\,cm containing Rb vapor. The cell temperature is kept at $85^{\circ}\,$C which corresponds to the maximum pick up probability for single Rb atoms. The latter was determined by recording resonant ionization yields of Rb$^+$ as a function of cell temperature. Further downstream, the beam of doped droplets intersects the focused laser beam  at right angle inside the detection volume of a commercial quadrupole mass spectrometer
(QMS). For beam analysis purposes, a Langmuir-Taylor (LT) detector is attached to the end of the beam line~\cite{Stienkemeier:2000}.

The femtosecond laser pulses are generated by a commercial mode-locked Ti:sapphire laser (Chameleon, Coherent) operating at a repetition rate of  80\,MHz. The average power output at the used wave length around $\lambda=773\,$nm amounts to about 2.5\,W. The pulses have a duration of about 200\,fs and a full spectral bandwidth at half maximum (FWHM) of 75\,cm$^{-1}$. Pairs of time delayed pulses with equal intensity are generated by a Mach-Zehnder interferometer. The new part in the present experiment is the addition of a femtosecond pulse shaper which is inserted into one arm of the interferometer. The pulse shaper consists of a liquid crystal based spatial light modulator with 640\,pixels (Jenoptik SLM S640d) placed in the Fourier plane of a $4f$-zero-dispersion compressor~\cite{Wefers:1995}.

\begin{figure}
\begin{center}{
\includegraphics[width=0.45\textwidth]{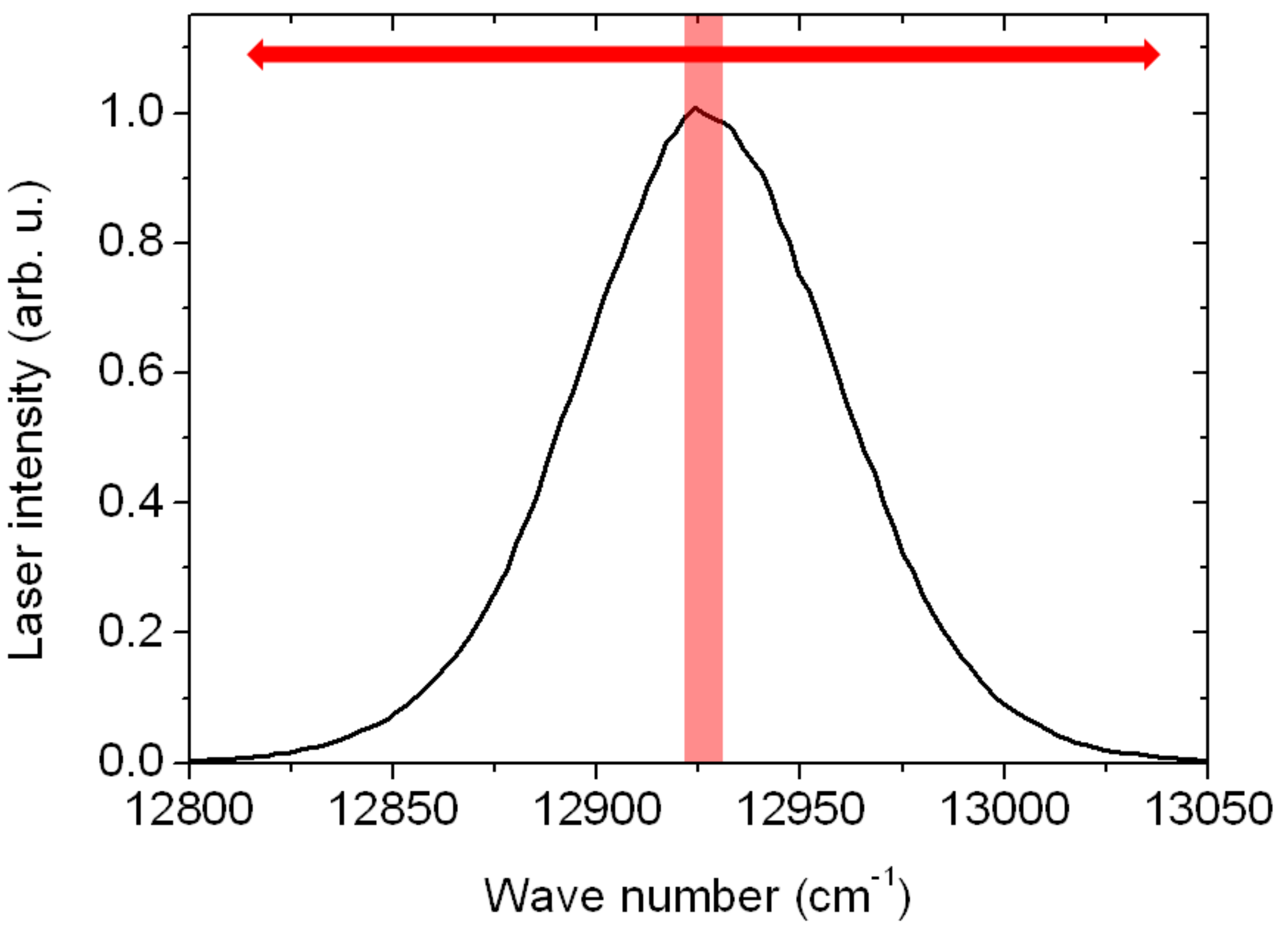}}
\caption{Measured spectrum of the femtosecond pump pulses (solid line). The transmission function of the pulse shaper is represented by the vertical shaded band which is tuned over the entire bandwidth of the femtosecond laser as indicated by the horizontal arrow.}
\label{fig:pulsespectrum}
\end{center}
\end{figure}
In the presented experiments we apply pure amplitude shaping with a square transmission function which has a width of 7 pixels which corresponds to a FWHM bandwidth of the modulated pulses of 9.1\,cm$^{-1}$. Fig.~\ref{fig:pulsespectrum} displays the measured spectral profile of the unshaped femtosecond laser pulses (solid line). The spectral transmission function of the amplitude shaper which is applied to the probe pulses is illustrated by the shaded vertical band. The horizontal arrow indicates the tuning range of the transmission function over the entire laser band width. As a result of the spectral narrowing, the shaped pulses are temporally stretched to about 3.4\,ps (FWHM). Due to the finite transmission of the gaps between active pixels, about 4\,\% of the input light intensity is transmitted even when all pixels are switched to zero transmission. This results in an additional background level in the probe pulse spectrum (not shown), which generates a constant offset in the yield of photoions which is subtracted from the measured data. Photoionization spectra are recorded by tuning the transmission window over the laser spectrum while measuring the yield of RbHe$^+$ photoions at various settings of the optical delay between unshaped pump and amplitude shaped probe pulses. Behind the interferometer, the shaped and unshaped laser beams are recombined with parallel polarizations and focused into the doped He droplet beam using a lens of 150\,mm focal length resulting in a 1/$e^2$-beam diameter of about $40\,\mu$m in the focus.

\section{Time-resolved R2PI spectra of $\mathbf{RbHe}$ exciplexes}
\label{sec:spectra}
\begin{figure}
\begin{center}{
\includegraphics[width=0.48\textwidth]{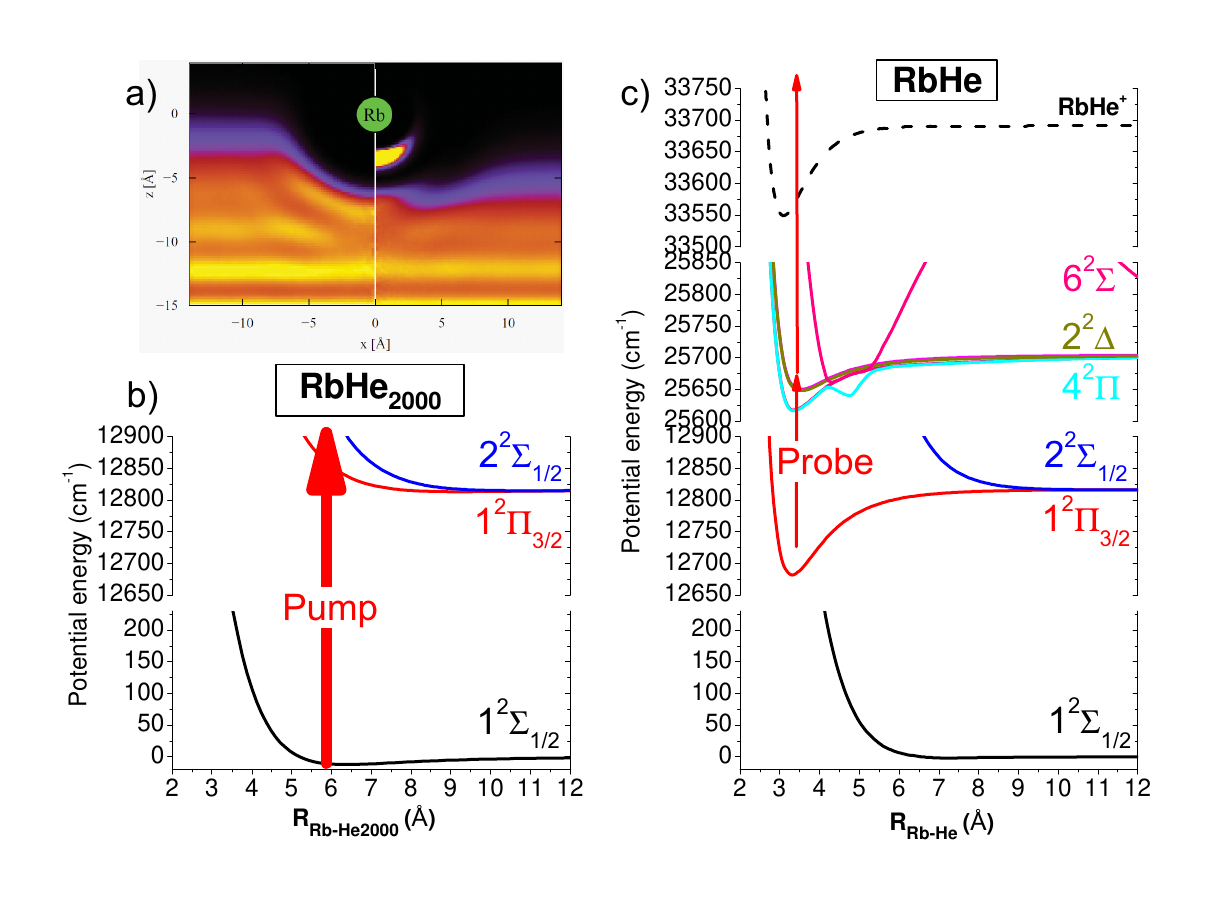}}
\caption{a) Density distribution of a $^4$He film with respect to an adsorbed Rb atom before (left half) and after (right half) the transition
from the electronic ground state $\Sigma_{1/2}$ to the excited $\Sigma_{3/2}$ state. The bright spot below the Rb$^*$ atom indicates the attachment of a He atom. Reprinted with permission from J.~Chem.~Phys.~\textbf{134}, 024316 (2011). Copyright 2011, American Institute of Physics. b)-c) Potential energy curves relevant to the RbHe formation and detection schemes. b) Pseudo-diatomic potentials of the Rb$^*$He$_N$ complex based on those of Ref.~\cite{Callegari:2011}. c) RbHe pair potentials based on those of Ref.~\cite{Pascale:1983}. The arrows indicate excitation by the fs pump pulse, followed by two-photon ionization by the amplitude-shaped probe pulse.}
\label{fig:levelscheme}
\end{center}
\end{figure}
The spectral lines of Ak atoms attached to He droplets are considerably shifted and broadened by up to $\lesssim 1000\,$cm$^{-1}$ compared to the free atomic lines due to the mostly repulsive interaction between the excited dopant Ak$^*$ and the He droplet in the Ak$^*$He$_N$ complex. The spectra are well reproduced by a pseudo-diatomic model (PDM) where the Ak dopant constitutes one atom and the whole He droplet the other. In this picture, which does not account for the internal degrees of freedom of the He droplet, upon electronic excitation the weakly bound AkHe$_N$ ground state undergoes mostly bound-free transitions to the repulsive short-range part of the excited PDM potentials. Thus, from the model it follows naturally that excited Ak$^*$ atoms mostly desorb off the He surface. Known exceptions are Rb and Cs on He droplets excited into their lowest PDM $1^2\Pi_{1/2}$ electronic states, which feature shallow potential wells at large distance from the droplet surface~\cite{Auboeck:2008,Theisen:2011}. Note that for Rb$^*$He in the $1^2\Pi_{3/2}$ state, Leino et al. predicted a surface-bound equilibrium configuration with a binding energy of 9.5\,cm$^{-1}$, where the Rb$^*$He molecular axis is oriented perpendicular to the droplet surface and the He atom is located between the Rb$^*$ atom and the surface (see Fig.~\ref{fig:levelscheme} a))~\cite{Leino:2011}. While the He surface forms a 4\,\AA~deep dimple in the vicinity of an attached Rb atom (left half of Fig.~\ref{fig:levelscheme} a)), the He dimple is much less pronounced in the case of the Rb$^*$He adduct (right half).

The relevant PDM potential curves obtained by modifying those of Ref.~\cite{Callegari:2011} in order to include spin-orbit coupling following the procedure described in Refs.~\cite{Reho2:2000,Bruehl:2001} are shown in Fig.~\ref{fig:levelscheme} b). Note that the distance $R_{Rb-He_{2000}}$ on the bottom axis specifies the distance between the Rb atom and the He surface region where the He density has dropped to half the bulk density. The corresponding potentials for the free RbHe diatom are shown on the right hand side. These potential curves are based on ab initio potentials modified to account for spin-orbit coupling as for the RbHe$_N$ potentials. The $1^2\Pi_{3/2}$ state supports 5 bound vibrational states~\cite{Mudrich:2008}.

The arrows in Fig.~\ref{fig:levelscheme} b) and c) indicate the transitions induced by pump and probe laser pulses in our experiment. At the laser wave lengths used ($\sim 773\,$nm) the fs pump pulse projects the PDM ground state $1^2\Sigma_{1/2}$ wave function onto the mostly repulsive part of the PDM $1^2\Pi_{3/2}$ potential which is likely to induce dissociation of the Rb$^*$He$_N$ complex. According to a classical trajectory calculation based on the PDM potentials the Rb$^*$ atom moves away from the He surface by $1\,$\,\AA~in about 2.2\,ps, after which it propagates nearly uniformly in free space. Simultaneously, Rb$^*$He exciplex molecules are being formed and are subsequently ionized by two-photon resonant ionization by the ps probe pulse via the state manifold correlating to the $5d$ level of atomic Rb.

The time scale on which Rb$^*$He formation takes place can be estimated using a tunneling model originally proposed by Reho et al.~\cite{Reho2:2000}. It is based on constructing an effective potential curve for the reaction coordinate of the single He atom that is being extracted out of the He droplet into the attractive region of the Rb$^*$He potential. The formation time constant $\tau$ is then estimated semiclassically from the tunneling probability from large distance through the shallow barrier towards the inner potential well. When assuming a distance of the excited Rb$^*$ atom from the He surface of $d=9.26\,$\AA~according to the PDM potentials, this procedure gives a tunneling time $\tau = 42\,$ps. This value is smaller than the originally estimated one, $\tau=220\,$ps~\cite{Droppelmann:2004}, but significantly longer than the experimental value $\tau\approx 8.5\,$ps inferred from pump-probe measurements~\cite{Droppelmann:2004,Mudrich:2008}. Therefore, we argue that a different faster formation mechanism must be active: The direct photoassociation-like excitation of bound states of Rb$^*$He in the inner potential well. This concept has been successfully applied to rationalizing the occurrence or absence of Rb$^*$He exciplexes upon excitation of higher-lying electronic levels~\cite{Fechner:2012}. Calculating Franck-Condon factors (FCF) of the bound-bound transition from the ground state of RbHe$_N$ into the modified $1^2\Pi_{3/2}$ pair potential of Rb$^*$He including the He extraction energy yields largest transition probability for the population of the highest bound state $1^2\Pi_{3/2}(v=4)$. Fast rearrangement of the He dimple environment following the excitation subsequently leads to the stabilization of Rb$^*$He in lower vibrational levels by energy relaxation into the He droplet. Quantum interference spectroscopy has revealed vibrational redistribution of population into lower lying vibrational states within about 15\,ps~\cite{Mudrich:2008}. Depending on the duration in which the Rb$^*$He remains attached to the droplet surface, further vibrational relaxation may take place, as it was observed with Rb$_2$ attached to He droplets excited to the lowest excited triplet state $1^3\Sigma_g^+$~\cite{Gruner:2011}.

\begin{figure}
\begin{center}{
\includegraphics[width=0.48\textwidth]{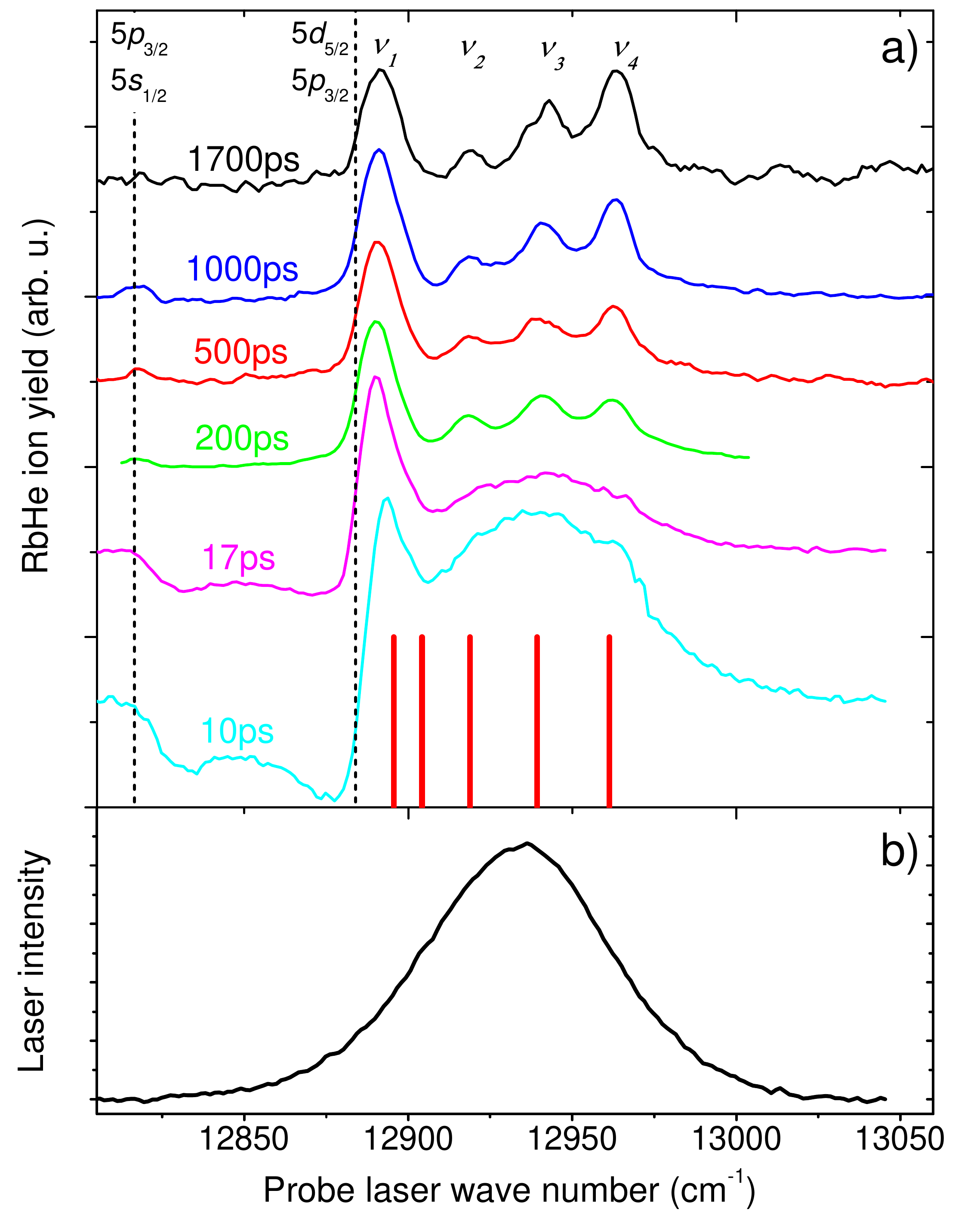}}
\caption{a) Measured RbHe$^+$ photoionization spectra obtained using a femtosecond pump pulse at the center frequency 12940\,cm$^{-1}$ and tunable amplitude-shaped probe pulses for various pump-probe delay times as indicated. The stick spectrum at the bottom depicts frequencies inferred from earlier quantum beam measurements. b) Spectrum of the unshaped femtosecond laser.}
\label{fig:spectra}
\end{center}
\end{figure}
The goal of the present study is to directly measure the vibrational spectrum of RbHe in the $1^2\Pi_{3/2}$ state and to follow in real time the populations of the individual vibrational levels. To this end, the narrow-band ps probe pulse is scanned in the entire spectral region covered by the fs laser at various values of the pump-probe delay. Typical RbHe$^+$ ion yield spectra are depicted in Fig.~\ref{fig:spectra}~a). Fig.~\ref{fig:spectra}~b) shows the spectral profile of the fs laser which defines the tuning range for the shaped probe pulses. The vertical dashed lines indicate the atomic $5s\rightarrow 5p_{3/2}$ and $5p_{3/2}\rightarrow 5d_{5/2}$ transition frequencies. Within the bandwidth of the laser profile, 4 peaks are clearly visible with increasing contrast as the delay time is stepped up from 10 to 1700\,ps. Since the peak around 12900\,cm$^{-1}$ is asymmetrically broadened towards higher wave numbers it is modeled by a sum of two gaussian functions which gives center wave numbers $\nu=12891(1)$ and $12900(1)$\,cm$^{-1}$, respectively. The latter wave number component is weak and quickly drops in amplitude within a few tens of ps and is therefore disregarded in the further analysis of the delay time dependent spectra. Thus, from gaussian fits we extract peak positions $\nu_1=12891(1)$, $\nu_2=12919(1)$, $\nu_3=12941(1)$, $\nu_4=12963(1)$\,cm$^{-1}$.
The undershooting of the ion signal between $12820$ and $12880$\,cm$^{-1}$ at short delay times is most likely due to an unrelated nonlinear photophysical effect or an experimental artifact. The same measurements were repeated for different settings of the central wave number of the fs laser in the range 12870-12990\,cm$^{-1}$. While the peak positions remain unchanged, the relative amplitudes vary according to the spectral intensity of the probe pulses at the particular peak positions. At wave numbers $<$12890\,cm$^{-1}$, two additional small peaks appear at $12855$ and $12875$\,cm$^{-1}$ at delay times $\gtrsim100\,$ps (not shown).

The stick spectrum shown at the bottom of Fig.~\ref{fig:spectra}~a) represents the frequencies inferred from earlier quantum interference measurements~\cite{Mudrich:2008} when assigning the observed beat spectrum to beats between transitions frequencies $\nu_v$, $v=0$-$3$ and $\nu_4$ instead of beats between adjacent $\nu_v$, $\nu_{v+1}$, as originally assumed. Here, $\nu_v$ stands for the transition frequencies into the vibrational levels $v$ of Rb$^*$He. The reasonably good agreement with the measured spectra suggests that the peaks are related to transitions to vibrational levels of Rb$^*$He. Note that high-contrast quantum interference oscillations are observed even with shaped probe pulses when scanning the delay time in steps $\lesssim1\,$fs at fixed probe pulse wave number. The interference contrast at different spectral positions of the probe pulse roughly matches the peak amplitudes shown in  Fig.~\ref{fig:spectra}~a) where quantum interferences are averaged out.

\begin{figure}
\begin{center}{
\includegraphics[width=0.42\textwidth]{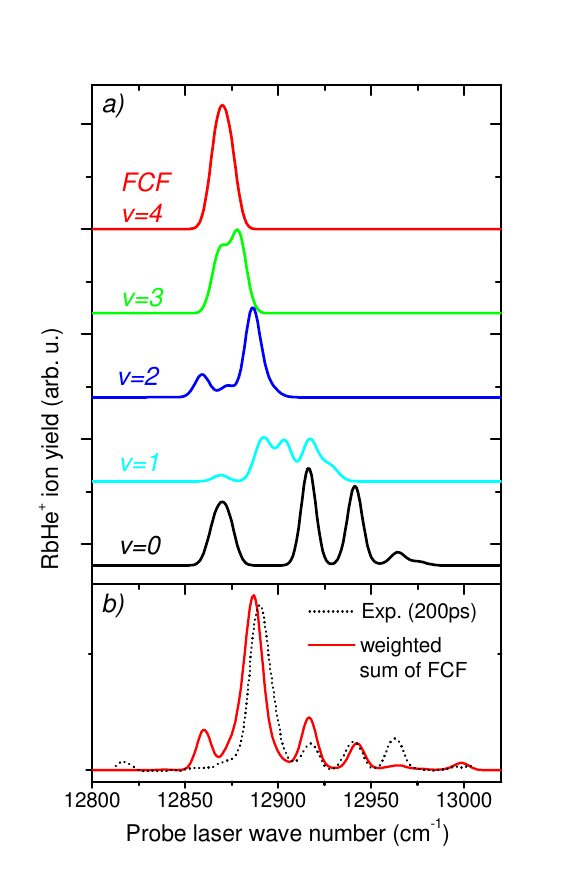}}
\caption{a) Calculated Franck-Condon factors for transitions from the indicated vibrational levels of the Rb$^*$He exciplex potential into all potential curves correlating to the $5d$ excitation of Rb, artificially broadened to FWHM line widths of $9$\,cm$^{-1}$. b) Weighted sum of the profiles shown in a) in comparison with the experimental data measured at a delay time of 200\,ps.}
\label{fig:FCF}
\end{center}
\end{figure}
Unfortunately, the assignment of the spectral features to vibronic transitions is not straight forward as in linear absorption spectroscopy. Since the measured RbHe$^+$ ion yield results from resonance-enhanced two-photon ionization (R2PI) with several contributing intermediate potential curves correlating to the $5d$ atomic level of Rb, namely $^2\Sigma_{1/2}$, $^2\Pi_{3/2,\,1/2}$, $^2\Delta_{3/2,\,5/2}$, the corresponding FCF spectra have to be considered. To this end, the spectral contribution of every individual $v$-level of the $1^2\Pi_{3/2}$-state is calculated by evaluating the FCF for all transitions into the $v$-levels of the Rb$^*$He states correlating to $5d$. The resulting stick spectra are artificially broadened by folding with a gaussian function with a width of $9$\,cm$^{-1}$ to account for the bandwidth of the shaped pulses and are shown in Fig.~\ref{fig:FCF} a). Fig.~\ref{fig:FCF} b) depicts the experimental data in comparison with a model curve obtained as the weighted sum of all spectral contributions shown in Fig.~\ref{fig:FCF} a) with weighting factors $p_0=0.15,\,p_1=0.05,\,p_2=0.80,\,p_3=0,\,p_4=0$ to best fit the experimental data recorded at 200\,ps delay (dashed line). The model curve clearly deviates from the experimental data in particular on the low and high wave number edges but the essential structure of the spectrum is approximately reproduced. Note that the accuracy of the potential curves is not well defined in particular as far as higher levels of excitation are concerned (compare for instance the Rb$^*$He potentials given in Ref.~\cite{Pascale:1983,Hirano:2003,Chattopadhyay:2012}). With this model the highest peak at $\nu_1$ is mainly associated with the population $p_2$ of level $v=2$ and peaks $\nu_2$-$\nu_4$ are best reproduced by the ionization spectrum of the vibrational ground state $v=0$. The peaks at 12855 and 12875\,cm$^{-1}$ observed at lower laser wave numbers can be attributed to the populations in levels $v=3,4$ and $v=2$, respectively.

\begin{figure}
\begin{center}{
\includegraphics[width=0.45\textwidth]{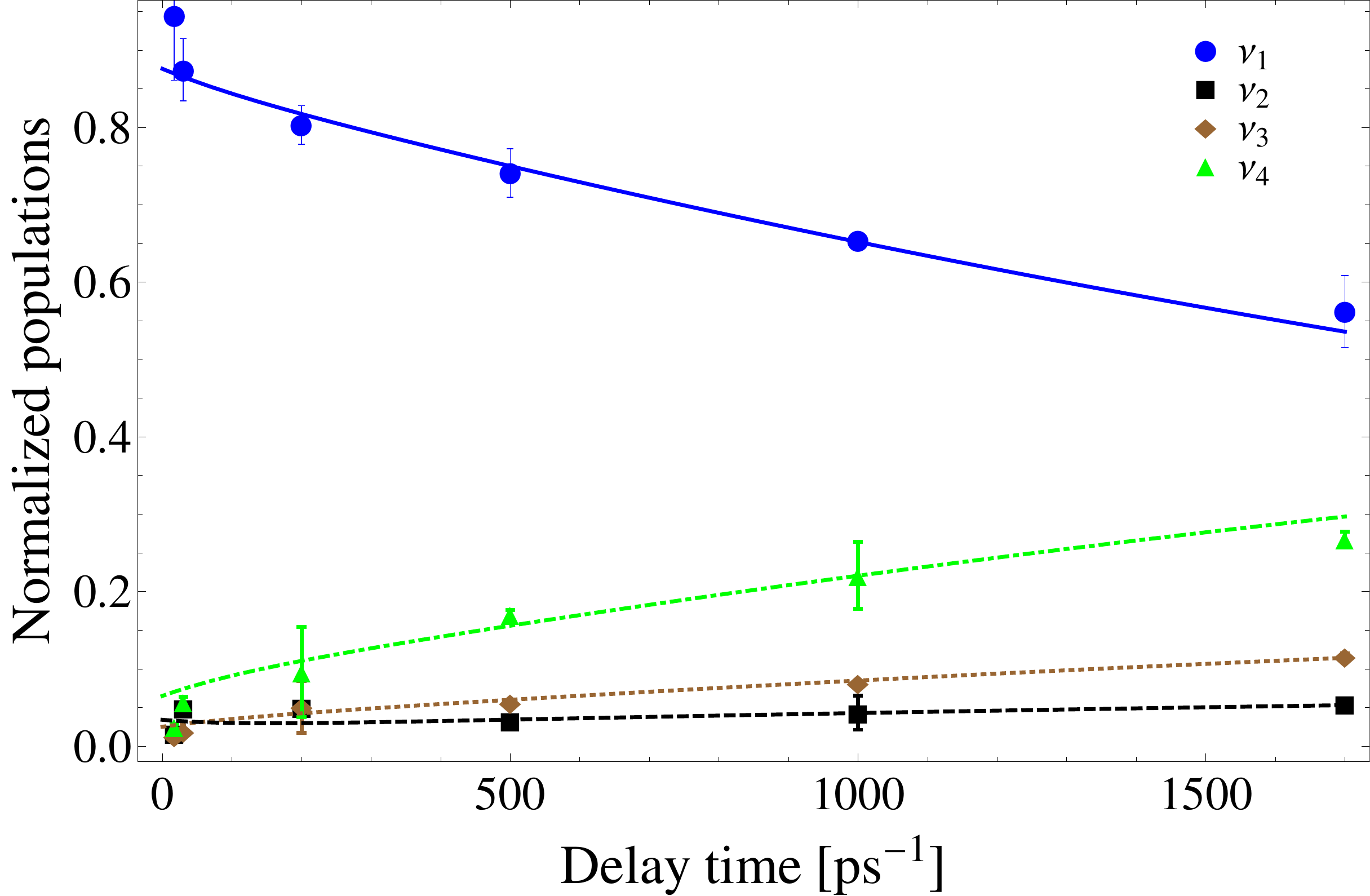}}
\caption{Peak amplitudes of the four dominant spectral features $\nu_{1-4}$ in the probe pulse spectra measured at different delay times obtained from gaussian fits (symbols). The solid lines represent the result of fitting the rate equation model~(\ref{eq:rates}) to the experimental data.}
\label{fig:dynamics}
\end{center}
\end{figure}

Finally, the time evolution of the spectra is analyzed by simultaneously fitting the model curve to all measured spectra. The varying intensity of the probe pulse as a function of the detuning is taken into account by linearly scaling the relative amplitudes of the spectral components according to the laser spectral profile. The weighting factors of the individual $v$-level contributions, which directly relate to the corresponding populations $p_v$ of $v$-levels, are varied according to a simple rate equation model that accounts only for sequential vibrational relaxation for the sake of simplicity,
\begin{equation}
\label{eq:rates}
\{\dot{p}_v=\gamma_{v+1}p_{v+1}-\gamma_v p_v\},\,v=0,\dots,4.
\end{equation}
In this set of equations, $\gamma_v$ denotes the rate coefficients for a decay from level $v$ into the adjacent level $v-1$ and $p_5\equiv 0$ and $\gamma_0=0$. The $v$-level spectra are taken as the calculated FCF for all $v$ except for $v=0$, for which we admit a slightly modified FCF spectrum with relative amplitudes 0.1/0.25/0.65 of the peaks at $\nu_2/\nu_3/\nu_4$ for better agreement with the measured spectra. The best results of fitting simultaneously all spectra recorded at different delay times on the basis of model~(\ref{eq:rates}) are obtained for a distribution of initial populations $p_0^0=0.1$, $p_1^0=0.05$, and $p_2^0=0.85$, $p_3^0=0$, $p_4^0=0$, and for rate coefficients $\gamma_2=0.28\,$ns$^{-1}$, $\gamma_1\gtrsim10\,$ns$^{-1}$. The resulting fit curves are depicted together with the experimental delay time dependent peak amplitudes in Fig.~\ref{fig:dynamics} as solid lines and as symbols, respectively. Thus, the time evolution of the measured spectrum can be nicely reproduced by a simple relaxation model provided relaxation rates as low as $0.28\,$ns$^{-1}$ are assumed.

This implies that at least part of the Rb$^*$He exciplexes remain attached to the droplet surface for much longer times than estimated when considering vertical excitation into the $1^2\Pi_{3/2}$ potential in the PDM which induces direct dissociation.
However, from time-resolved quantum interference spectroscopy we inferred that fast vibrational stabilization in lower-lying $v$-levels $v=0-3$ of Rb$^*$He proceeds within about 15\,ps. This indicates that the dynamics proceeds on two distinct time scales following laser excitation: The local He environment of the excited Rb$^*$ atom rearranges within a few picoseconds owing to the competing repulsion of excited Rb$^*$ with respect to the whole droplet and the attractive pair interaction between Rb$^*$ and single He atoms. This fast dynamics seems to lead to a non-thermal distribution of populations of bound vibrational states of the Rb$^*$He molecule peaked at intermediate levels through dissipative coupling to the He droplet. However, as a surprising outcome of this work, there is a second relaxation mechanism on a much longer time scale. We indeed find that the vibrational population continues to relax towards the ground state even after delay times as long as 1.7\,ns.

As an alternative mechanism that may induce the observed slow relaxation we have considered the fast desorption of Rb$^*$He exciplexes followed by inelastic collisions with evaporating He atoms. The He-Rb$^*$He collision rate $\gamma_{coll}$ per Rb$^*$He is estimated based on the evaporation rate of He clusters~\cite{Brixner:2000} and the Langevin capture cross section $\sigma=3\pi\sqrt[3]{C_6/(4E_{coll})}$ for vibrationally inelastic collisions~\cite{Levine:2005}. The long-range dispersion coefficient $C_6$ for the Rb$^*$He-He interaction is approximated by the one for the Rb$^*$-He interaction, $C_6\approx 70\,$a.\,u.~\cite{Zhang:2007}. The He evaporation rate amounts to $\gamma_{evap}\sim 10^{10}\,$s$^{-1}$ at times $0\lesssim t\lesssim 100\,$ps after laser excitation and falls off sharply for longer times. When assuming isotropic evaporation of He atoms into the full solid angle, we arrive at a scattering rate $\gamma_{coll}\lesssim 5\times 10^7\,$s$^{-1}$ at all relevant times, which is by far too low to account for the measured relaxation rates. Only when admitting He atoms to be ejected into a narrow solid angle cone pointing towards the desorbing Rb$^*$He with an opening angle $\lesssim 5^\circ$ the rate $\gamma_{coll}$ reaches the measured values $\gamma_v\sim 10^8\,$s$^{-1}$ at times $t\sim 1\,$ns. As such a strong anisotropy in the evaporation appears highly implausible, we tend to rule out this scenario.

\section{Conclusion}
\label{sec:conclusion}
The formation and relaxation process of Rb$^*$He exciplexes initiated by the excitation of Rb atoms on the surface of He nanodroplets has been studied using femtosecond pump and picosecond probe spectroscopy. The picosecond probe pulse is obtained by amplitude shaping of a femtosecond pulse and is individually tuned within the bandwidth of the femtosecond laser at variable delay times. This allows to follow the time evolution of the ionization spectrum
of Rb$^*$He on the time scale of hundreds of picoseconds up to nanoseconds. By fitting a model of the spectra based on the Franck-Condon-factors for vibronic transitions in free Rb$^*$He we find the populations of Rb$^*$He vibrational states to be initially set to intermediate levels by strong transient coupling to the droplet environment and to relax towards the ground state even after delay times as long as 1.7\,ns.

A similar behavior of fast redistribution of population into different modes followed by slow relaxation has been observed before with alkali dimers and trimers~\cite{Gruner:2011,Giese:2011,Reho:2001}. Thus, we conclude that Rb$^*$He exciplexes formed in the $1^2\Pi_{3/2}$ state most probably remain attached to He droplets where they are subjected to weak dissipative interaction with the He environment which acts as a bath. This interpretation is supported by recent ion imaging measurements that revealed qualitatively different velocity distributions of desorbing sodium atoms as compared to those of Na$^*$He exciplexes formed in the lowest excited state of Na~\cite{LoginovPhD:2008}. Further experiments and detailed theoretical calculations, as recently performed for the repulsive $^2\Sigma_{1/2}$ states of lithium and sodium~\cite{Hernando:2012}, are needed to gain the full understanding of the quantum dynamics of the alkali-helium exciplex formation process on He nanodroplets.

\begin{acknowledgments}
We gratefully acknowledge support by DFG.
\end{acknowledgments}


\end{document}